\documentclass{article}

\usepackage[preprint]{neurips_2024}

\usepackage[utf8]{inputenc} 
\usepackage[T1]{fontenc}    
\usepackage{hyperref}       
\usepackage{url}            
\usepackage{booktabs}       
\usepackage{amsfonts}       
\usepackage{nicefrac}       
\usepackage{microtype}      
\usepackage{amsmath}
\usepackage{amsthm}
\usepackage{graphicx}
\usepackage{pifont}
\usepackage{bbm}
\usepackage{bm}
\usepackage{caption}
\usepackage[dvipsnames]{xcolor}
\usepackage{arydshln}
\usepackage{amsmath}
\usepackage{amssymb}
\usepackage{mathtools}
\usepackage{amsthm}
\usepackage{amssymb}
\usepackage{colortbl}

\newtheorem{theorem}{Theorem}[section]

\theoremstyle{definition}

\usepackage{multirow}
\usepackage{adjustbox}
\usepackage[linesnumbered,ruled]{algorithm2e}

\title{Fast Training-free Perceptual Image Compression}

\author{%
  \textbf{Ziran Zhu*\textsuperscript{1,3,8}}, \textbf{Tongda Xu*\textsuperscript{1,2}},
  \textbf{Minye Huang\textsuperscript{6}}, \textbf{Dailan He\textsuperscript{5}},\textbf{Xingtong Ge\textsuperscript{4}},\\
  \textbf{Xinjie Zhang\textsuperscript{7}},\textbf{Ling Li   \textsuperscript{1,3\dag} },
  \textbf{Yan Wang  \textsuperscript{1,2\dag} }\\
  \textsuperscript{1}Institute for AI Industry Research, Tsinghua University \\
  \textsuperscript{2}Department of Computer Science and Technology, Tsinghua University\\
  \textsuperscript{3}Institute of Software, Chinese Academy of Sciences \\
  \textsuperscript{4}SenseTime Research\\
  \textsuperscript{5}The Chinese University of Hong Kong \\
  \textsuperscript{6}Harbin Institute of Technology \\
  \textsuperscript{7}Hong Kong University of Science and Technology \\
  \textsuperscript{8}University of Chinese Academy of Sciences\\
}

\begin{document}

\maketitle

\begin{abstract}
Training-free perceptual image codec adopt pre-trained unconditional generative model during decoding to avoid training new conditional generative model. However, they heavily rely on diffusion inversion or sample communication, which take 1 min to intractable amount of time to decode a single image. In this paper, we propose a training-free algorithm that improves the perceptual quality of any existing codec with theoretical guarantee. We further propose different implementations for optimal perceptual quality when decoding time budget is $\approx 0.1$s, $0.1-10$s and $\ge 10$s. Our approach: 1). improves the decoding time of training-free codec from 1 min to $0.1-10$s with comparable perceptual quality. 2). can be applied to non-differentiable codec such as VTM. 3). can be used to improve previous perceptual codecs, such as MS-ILLM. 4). can easily achieve perception-distortion trade-off. Empirically, we show that our approach successfully improves the perceptual quality of ELIC, VTM and MS-ILLM with fast decoding. Our approach achieves comparable FID to previous training-free codec with significantly less decoding time. And our approach still outperforms previous conditional generative model based codecs such as HiFiC and MS-ILLM in terms of FID. The source code is provided in the supplementary material.
\end{abstract}
\section{Introduction}
\begin{table}[h]
\begin{minipage}[t]{0.6\linewidth}
\vspace{0pt}
\centering
\includegraphics[width=\linewidth]{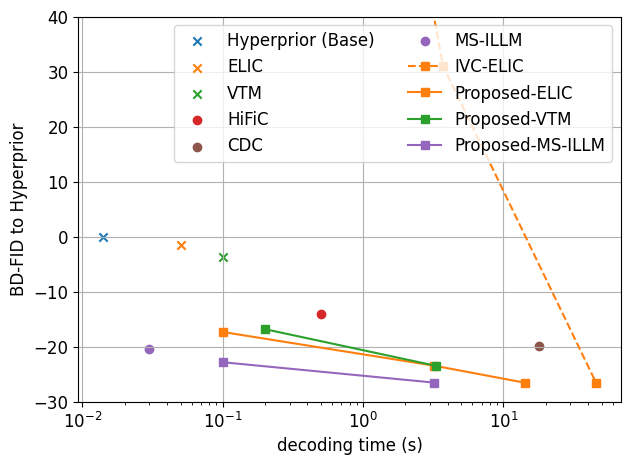}
\captionof{figure}{BD-FID vs decoding time of different methods. IVC and Proposed are training-free.}
\label{fig:cov}
\end{minipage}
\hfill
\begin{minipage}[t]{0.4\linewidth}
\centering
\vspace{0pt}
\captionof{table}{Temporal complexity comparison of different methods.}
\label{tab:rlow}
\resizebox{\linewidth}{!}{
\begin{tabular}{@{}lll@{}}
\toprule
                & Train  & Decode \\ \midrule
HiFiC, MS-ILLM   & 1 week & $\approx 0.1$s   \\
IVC             & 0      & 60s  \\
Proposed (fast) & 0      & 0.1s   \\
Proposed (med) & 0      & 0.1-10s   \\
Proposed (slow) & 0      & 10s   \\\bottomrule
\end{tabular}
}
\end{minipage}\hfill
\end{table}
Perceptual image codecs aim to compress source images into low bitrate with perceptually lossless reconstruction, \textit{i.e.,} people can not distinguish the source images and decoded images. Early works of perceptual image codec train a conditional generative model \citep{tschannen2018deep,agustsson2019generative} to achieve perceptual quality. This practice is later justified by theory of rate-distortion-perception trade-off \citep{blau2018perception,blau2019rethinking}. After that, the majority of perceptual codec follow the conditional generative model paradigm. They either train conditional generative adversial network \citep{Mentzer2020HighFidelityGI,Agustsson2022MultiRealismIC,he2022po,muckley2023improving} or conditional diffusion model \citep{yang2022lossy,goose2023neural,hoogeboom2023high,careil2023towards} to achieve perceptual quality.

On the other hand, training-free perceptual codec has gained great attention recently, due to the increasing cost to train a state-of-the-art diffusion model. \citet{Theis2022LossyCW} are the first to use pre-trained unconditional diffusion model for perceptual codec. They propose to use correlation communication \citep{harsha2007communication,li2018strong} to encode the trajectory of diffusion process, which is intractable so no actual codec is implemented. \citet{xu2024idempotence} improve the decoding complexity from intractable to approximately 1 min, by adopting diffusion inversion \citep{chung2022diffusion}. However, a decoding time of 1 min is still way too high for practical use.

In this paper, we propose a fast training-free perceptual codec. More specifically, we propose to run the forward diffusion step on decoded image and reverse the diffusion directly. We theoretically show that this approach guarantees to reduce the KL divergence between the source image and compressed image. Our approach has several advantages:
\begin{itemize}
    \item As no correlation communication \citep{harsha2007communication} or diffusion inversion \citep{Chung2022DiffusionPS} is involved, our training-free approach decodes in 0.1-10s.
    \item Unlike previous training-free codec \citep{xu2024idempotence} that requires the base codec to be differentiable, our approach works for non-differentiable codec such as VTM.
    \item Our approach can also be applied to further improve the perceptual quality of a pre-trained perceptual codec, such as MS-ILLM \citep{muckley2023improving}.
    \item When the base codec is MSE optimized, our approach can achieve perception-distortion trade-off.
\end{itemize}
    
Compared with previous conditional generative codec, our approach achieves better perceptual quality in terms of FID. Compared with previous training-free codec, our approach reduces decoding time from 1 min to 0.1-10s with marginal increase in FID.

\section{Preliminaries}
\subsection{Perceptual Image Compression} 
Denote the source image as $X_0$, the encoder as $f(.)$, the bitstream as $Y=f(X_0)$, the decoder as $g(.)$ and the decompressed image as $\hat{X} = g(Y)$, the perceptual image codec constrains the divergence between the source and decompressed image. More specifically, perceptual codec requires:
\begin{gather}
    d(p(X_0),p(\hat{X})) = 0, i.e, p(X_0) = p(\hat{X}). \label{eq:piq}
\end{gather}
\textbf{Conditional Generative Perceptual Codec.} The majority of perceptual codecs achieve the constraint by training a conditional decoder \citep{Mentzer2020HighFidelityGI,hoogeboom2023high}. This conditional decoder can be any conditional generative model, such as GAN and diffusion. More specifically, they train the decoder to match the posterior of original image given bitstream:
\begin{gather}
    \hat{X} = g(Y) \sim p(X|Y).
\end{gather}
\citet{blau2018perception,blau2019rethinking} show that perceptual codec can be achieved at the cost of doubling mean-square-error (MSE):
\begin{gather}
    \mathbb{E}[||\hat{X} - X_0||^2] \le 2\mathbb{E}[||\mathbb{E}[X_0|Y]-X_0||^2]
\end{gather}
\citet{yan2022optimally} further show that when encoder is deterministic, this bound is tight, \textit{i.e.,} at least doubling MSE is required.

\textbf{Training-free Perceptual Codec.} Recent state-of-the-art generative models, such as diffusion model \citep{ho2020denoising,Dhariwal2021DiffusionMB}, are hard to train. To tackle this challenge, \citet{Theis2022LossyCW} are the first to propose training-free perceptual codec using diffusion model. More specifically, they encode the per step sample of diffusion process by sample communication \citep{harsha2007communication,li2018strong}. However, for practical size image their approach is intractable. 

\citet{xu2024idempotence} reduce the complexity of \citet{Theis2022LossyCW} by using diffusion inversion \citep{Chung2022DiffusionPS} instead of sample communication. More specifically, they solve the optimization problem below with diffusion prior. However, their approach still require 1 min of decoding time, which is impractical.
\subsection{Diffusion Models and Stochastic Differentiable Equations} 
Diffusion model is a type of generative model whose inference and generation process are both $T$ steps Gaussian Markov chain. Roughly speaking, there are two types of diffusion model: variance preserving (VP) and variance exploding (VE) diffusion. In this paper, we use notation of VE diffusion. And we refer readers interested in VP diffusion to \citep{ho2020denoising}. More specifically, given fixed sequence of $\sigma_1^2 <...< \sigma_t^2<...< \sigma_T^2$, the inference and generation process of VE diffusion are:
\begin{gather}
    X_t = X_{t-1} + \mathcal{N}(0,(\sigma_t^2 - \sigma_{t-1}^2)I), \\
    X_{t-1} = X_t + (\sigma_t^2 - \sigma_{t-1}^2) s_{\theta}(t,X_t) + \mathcal{N}(0, (\sigma_t^2 - \sigma_{t-1}^2)I),\label{eq:rdiff}
\end{gather}
where $s_{\theta}(t,X_t)$ is a neural network parameterized by $\theta$ approximating the score function $\nabla \log p(X_t)$ learned by conditional score matching \citep{vincent2011connection}.

\citet{song2020score} show that the above discrete time diffusion process is equivalent to the discretization of the forward and backward stochastic differential equation (SDE) \citep{anderson1982reverse}:
\begin{gather}
    \textrm{forward SDE: } dX_t = \sqrt{\frac{d\sigma^2}{dt}}dB_t,\textrm{reverse SDE: } dX_t = -\frac{d \sigma^2_t}{d_t} s_{\theta}(t,X_t) dt + \sqrt{\frac{d \sigma^2_t}{d_t}}dB_t.\label{eq:rsde}
\end{gather}
And furthermore, there exists a probability flow ordinary differential equation (PF-ODE) that has the same marginal distribution as reverse SDE:
\begin{gather}
    \textrm{PF-ODE: } dX_t = -\frac{1}{2}\frac{d\sigma^2_t}{dt}s_{\theta}(t,X_t)dt.
    \label{eq:pfode}
\end{gather}
And therefore, we can generate samples from the learned diffusion by solving the above PF-ODE with existing ODE solvers. Another useful result is Tweedie's formula, which allows fast estimation of posterior mean at any time $t$:
\begin{gather}
    \mathbb{E}[X_0|X_t] = X_t + \sigma^2 s_{\theta}(t,X_t).\label{eq:tw}
\end{gather}

\section{Fast Training-Free Perceptual Image Compression}
\subsection{Perceptual Enhancement with Theoretical Guarantee}
We propose a simple yet effective way to enhance the perceptual quality of any codec by reducing the divergence between decoded image $\hat{X}$ and real image $X_0$. More specifically, we consider the following procedure on decoder side:
\begin{gather}
\textrm{Decoded: } \hat{X} = g(Y),\\
\textrm{Add noise: } \hat{X}_t = \hat{X} + \mathcal{N}(0, \sigma_t^2I),\\
\textrm{Denoise: } \hat{X}_{0|t} = \Phi(\hat{X}_t,t),
\end{gather}
where $\Phi(.)$ is either the reverse SDE solver or PF-ODE solver. The first "Add noise" step obviously reduces the divergence, as both distributions are convoluted with the same Gaussian Markov kernel. A very recent literature \citep{nie2024the} shows the second "Denoise" reduces KL-divergence when $\Phi(.)$ is SDE solver, and preserves KL-divergence when $\Phi(.)$ is ODE solver. Combining them together, our approach has theoretical guarantee:
\begin{theorem}
\label{thm:01} Our proposed approach improves the perceptual quality of codec. Formally, when $p(\hat{X}) \neq p(X_0)$, we have
\begin{gather}
D_{KL}(p(\hat{X}_{0|t})||p(X_0)) < D_{KL}(p(\hat{X})||p(X_0)).
\end{gather}
Furthermore, the higher the $\sigma_t^2$ is, the closer the resulting KL divergence is, \textit{i.e.,}
\begin{gather}
    \forall s<t, D_{KL}(p(X_{0|t})||p(X_0)) < D_{KL}(p(X_{0|s}||p(X_0)).
\end{gather}
\end{theorem}
In fact, similar approach is already used in diffusion based image editing \citep{meng2021sdedit}. In those literature, the noise $\sigma_t^2$ is selected empirically. The higher $\sigma_t^2$ is, the better the perceptual quality is. However, the result also differs more from the original image.

Previous work of image editing \citep{meng2021sdedit} selects $\sigma_t^2$ empirically. For image compression, the selection of $\sigma_t^2$ can be guided by distortion-perception trade-off \citep{blau2018perception,blau2019rethinking}.
\begin{gather}
\textrm{Noise Selection: } \sigma_t^{2*} = \max \sigma_t^2 \textrm{, s.t.} \mathbb{E}[||\hat{X}_{0|t}-X_0||^2] \le 2\mathbb{E}||\Delta(\mathbb{E}[X_0|Y],X_0||) 
\end{gather}
Empirically, we find that when $\sigma_t^2$ is large enough such that $\mathbb{E}[||\hat{X}_{0|t}-X||] \approx 2\mathbb{E}[||\mathbb{E}[X_0|Y]-X_0||^2]$, the FID curve is also flat. So usually the best $\sigma_t^2$ is the one that doubles MSE (See Fig.~\ref{fig:pd}).

\subsection{Implementation for Speed-Performance Trade-off}
By choosing different implementation of the proposed approach, we can achieve a wide-range trade-off between decoding time and perceptual quality, from $0.1$s to $\ge 10$s. Furthermore, all of them achieve better perceptual quality than previous training-free codec with same decoding time. In analogous to the "fast, medium, slow" preset in traditional codec \citep{merritt2006x264}, we also use "fast, medium, slow" to name the optimal implementation for different time constraint. Those implementations are determined empirically in Fig.~\ref{fig:pt}, and the results are shown in Tab.~\ref{tab:fms}.
\begin{table}[htb]
\caption{The decoding time and implementation of different presets.}
\centering
\begin{tabular}{@{}lll@{}}
\toprule
Preset & Decoding Time  & Implementation \\ \midrule
Fast   & $\approx 0.1$s & Distilled ODE solver with consistency model                               \\
Medium & $0.1-10$s      & ODE solver for $0.1-5$s, SDE solver for $5-10$s  \\
Slow   & $\ge 10$s      & SDE solver with diffusion inversion constraint \\ \bottomrule
\end{tabular}
\label{tab:fms}
\end{table}

\textbf{Fast Preset for $\approx 0.1$s Decoding with Consistency Model.}
One exciting part of the proposed approach is that as no diffusion inversion \citep{Chung2022DiffusionPS} is required, we can adopt one-step distilled PF-ODE as an approximation. More specifically, consistency model (CM) \citep{Song2023ConsistencyM} can solve PF-ODE from any $\sigma^t_t$ by a single neural function evaluation (NFE):
\begin{gather}
    g_{\theta}(t,X_t) \approx \Phi^{ODE}(t,X_t),
\end{gather}
where $g_{\theta}(t,X_t)$ is the consistency model neural network. This neural network has same structure as the score function estimator $s_{\theta}(t,X_t)$. Therefore, the inference time of consistency model is only $1/t$ of inference time of a $t$ step SDE/ODE solver. And by using CM, we can implement our proposed training-free codec with only $0.1$s, which is faster than previous training-free codecs \citep{Theis2022LossyCW,xu2024idempotence}. And this decoding time is even close to previous conditional generative codecs such as HiFiC \citep{Mentzer2020HighFidelityGI} and MS-ILLM \citep{muckley2023improving}.

\textbf{Medium Preset for $0.1-10$s Decoding with SDE/ODE Solvers.}
When the constraint on decoding time is beyond one step SDE/ODE solvers, the SDE/ODE solvers with several steps are good choices. The error of SDE/ODE solvers come from two sources: the approximation error and discretization error. The approximation error is the distance of score approximator $s_{\theta}(t.X_t)$ to true score. The discretization error is caused by discrete SDE/ODE solvers. When the step number of SDE/ODE solver is large, the approximation error dominates. When the step number of SDE/ODE solver is small, the discrtization error dominates. 

Previous literature has shown that SDE achieves better sample quality than ODE
\citep{deveney2023closing} in terms of approximation error. However, ODE solvers usually converge faster than SDE solvers when the step number is small \citep{song2020denoising,lu2022dpm}, and have smaller discrtization error. Thus theoretically, when the decoding time constraint is tighter, ODE solver is preferred. And when the decoding time constraint is less tight, SDE solver is preferred. In practice, we observe this phenomena as shown in Fig.~\ref{fig:pt}.

\textbf{Slow Preset  for $\ge 10$s Decoding with SDE and Diffusion Inversion Constraint.} When the decoding time budget is $\ge 10$s and the base codec is differentiable, we can add diffusion inversion constraint to SDE solvers. We consider the diffusion posterior sampling (DPS) constraint \citep{Chung2022DiffusionPS}. More specifically, we modify the SDE update from Eq.~\ref{eq:rdiff} to
\begin{gather}
    X_{t-1} = X_t + (\sigma_t^2 - \sigma_{t-1}^2) s_{\theta}(t,X_t) + \mathcal{N}(0, (\sigma_t^2 - \sigma_{t-1}^2)I),\\
    X_{t-1} = X_{t-1} - \zeta \nabla_{X_t} ||f(\mathbb{E}[X_0|X_t]) - Y||\label{eq:dps}.
\end{gather}
Eq.~\ref{eq:dps} is the additional constraint by DPS. $\mathbb{E}[X_0|X_t]$ is evaluated by Tweedie's formula in Eq.~\ref{eq:tw} and $\zeta$ is a hyper-parameter.

We can also understand this approach as initializing DPS with the noise perturbed decoded image $\hat{X}_t$, and run diffusion inversion subsequently from $\hat{X}_t$. This initialization is shown to make linear diffusion inversion faster \citep{chung2022come}, but has not been examined for non-linear diffusion inversion such as codec.

The diffusion inversion such as DPS \citep{Chung2022DiffusionPS} does not perform well when the step number is small. Further, its per-step complexity is higher than SDE/ODE solvers as in general we not only need the value of $s_{\theta}(t,X_t)$ but also need the gradient of $s_{\theta}(t,X_t)$. Empirically, we show that this approach has advantage over ODE solvers only when decoding time budget is more than $10$s in Fig.~\ref{fig:pt}.

\subsection{Perception-Distortion Trade-off}

Another feature of the proposed approach is that it achieves perception-distortion trade-off naturally when the base codec is MSE optimized. If we directly decode $\hat{X}$ without adding noise and denosing, the decoded image is $\mathbb{E}[X|Y]$, which achieves best MSE/PSNR possible. At that time we can treat it as adding a $\sigma_t^2 = 0$ to $\hat{X}$ and denoising. If we add noise $\sigma_t^{*2}$ so that the denoised image $\hat{X}_{0|t}$ has $-3$dB PSNR compared with $\hat{X}$, we achieve the best perception within the distortion lowerbound. A natural way to achieve perception-distortion trade-off is traversing $0<\sigma^2_t<\sigma_t^{*2}$. Then the denoised image will also traverse through the best PSNR to worst PSNR, and worst perception to best perception.

\subsection{Comparison to Previous Works}
Compared with previous works using conditional generative model \citep{Mentzer2020HighFidelityGI, muckley2023improving}, our approach does not require training a new model and has comparable decoding time. Besides, our approach can achieve perception-distortion trade-off easily, without complex $\beta$-conditional training \citep{agustsson2023multi,iwai2024controlling}.

Compared with previous training-free codec \citep{Theis2022LossyCW,xu2024idempotence}, our approach reduces the decoding complexity from more than $1$ min to $\approx 0.1$s. Furthermore, our approach does not require the base codec to be differentiable.

Our approach is universal to base codec with bitstream compatibility. More specifically, it can: 1). Turn a MSE optimized codec, such as ELIC \citep{he2022elic}, into a perceptual codec. 2).Turn a traditional codec, such as VTM, into a perceptual codec. 3). Improve the perceptual quality of a perceptual codec, such as MS-ILLM \citep{muckley2023improving}.

\section{Experiments}
\label{sec:exp}
\subsection{Experiment Setup} 

\textbf{Base Diffusion Models.} A pre-trained diffusion model and consistency model is needed by our method . We adopt \textbf{EDM} and \textbf{ADM} as base diffusion model. EDM is a more recent diffusion model with pre-trained CM available \citep{Song2023ConsistencyM}. ADM is the diffusion model used by previous training-free codec \citep{xu2024idempotence}. 

\textbf{Base Codec Models.} A base codec is needed in our method to produce the bitstream and decoded image $\hat{X}$. We select \textbf{ELIC} \citep{he2022elic}, \textbf{VTM} and \textbf{MS-ILLM} \citep{muckley2023improving} as base codec to run our approach. As we need to test our codec's ability to: 1). Turn a MSE optimized codec into a perceptual codec. 2). Turn a non-differentiable traditional codec into a perceptual codec. 3). Further improve the perceptual quality of a perceptual codec.

As we have stated previously, we select our $\sigma_t^2$ according to the MSE lowerbound induced by perception-distortion trade-off \citep{blau2018perception, blau2019rethinking}. ELIC and VTM are MSE optimized, and we think the MSE between decoded image and source is the optimal MSE that can be achieved. For MS-ILLM, we use the MSE of Hyper as optimal MSE, because MS-ILLM's base codec is approximately the same as Hyper.

\textbf{Dataset.} As our base diffusion models and CM are pre-trained on \textbf{LSUN Bedroom} and \textbf{ImageNet} dataset, we use the first 1000 image of LSUN Bedroom and Imagenet validation dataset as test dataset. Following previous works \citep{xu2024idempotence} in training-free perceptual codec, we scale the image's short edge into $256$ and crop image into a square. 

\textbf{Metrics.} For perceptual quality evaluation, we use \textbf{Fréchet Inception Distance (FID)} \citep{heusel2017gans}. On one hand, our definition of perceptual quality is divergence based \citep{blau2018perception}. And FID is the most common choice to measure the divergence of two image distribution. On the other hand, from statistics of CLIC 2022 competition, FID is most aligned with ELO score from real human test. Other metrics such as MS-SSIM correlates poorly to ELO score. For distortion and bitrate metric, we adopt \textbf{Peak-Signal-Noise-Ratio (PSNR)} and \textbf{bits-per-pixel (bpp)}. As codec has different bitrate, Bjontegaard (BD) metrics \citep{bjontegaard2001calculation} are used to compute BD-FID and BD-PSNR. They are the average FID and PSNR covering the range of bpp.

\textbf{Previous State-of-the-art Perceptual Codec.} For MSE optimized codec, we select \textbf{Hyperprior} \citep{Ball2018VariationalIC} and \textbf{ELIC} \citep{he2022elic}. For traditional codec, we select \textbf{VTM}. 
For conditional generative model based perceptual codec, we select \textbf{HiFiC} \citep{Mentzer2020HighFidelityGI}, \textbf{CDC} \citep{yang2022lossy}, \textbf{Po-ELIC} \citep{he2022po} and \textbf{MS-ILLM} \citep{muckley2023improving} as baseline methods. For training-free codec, we use \textbf{IVC} \citep{xu2024idempotence} as it is the only practical training-free perceptual codec. We acknowledge that there are other highly competitive conditional generative perceptual codec \citep{iwai2021fidelity,ma2021variable,Agustsson2022MultiRealismIC,goose2023neural,hoogeboom2023high,careil2023towards}. However, we can not include them for comparison as no inference code is provided.

\subsection{Main Results}
\begin{figure}[thb]
\centering
    \includegraphics[width=\linewidth]{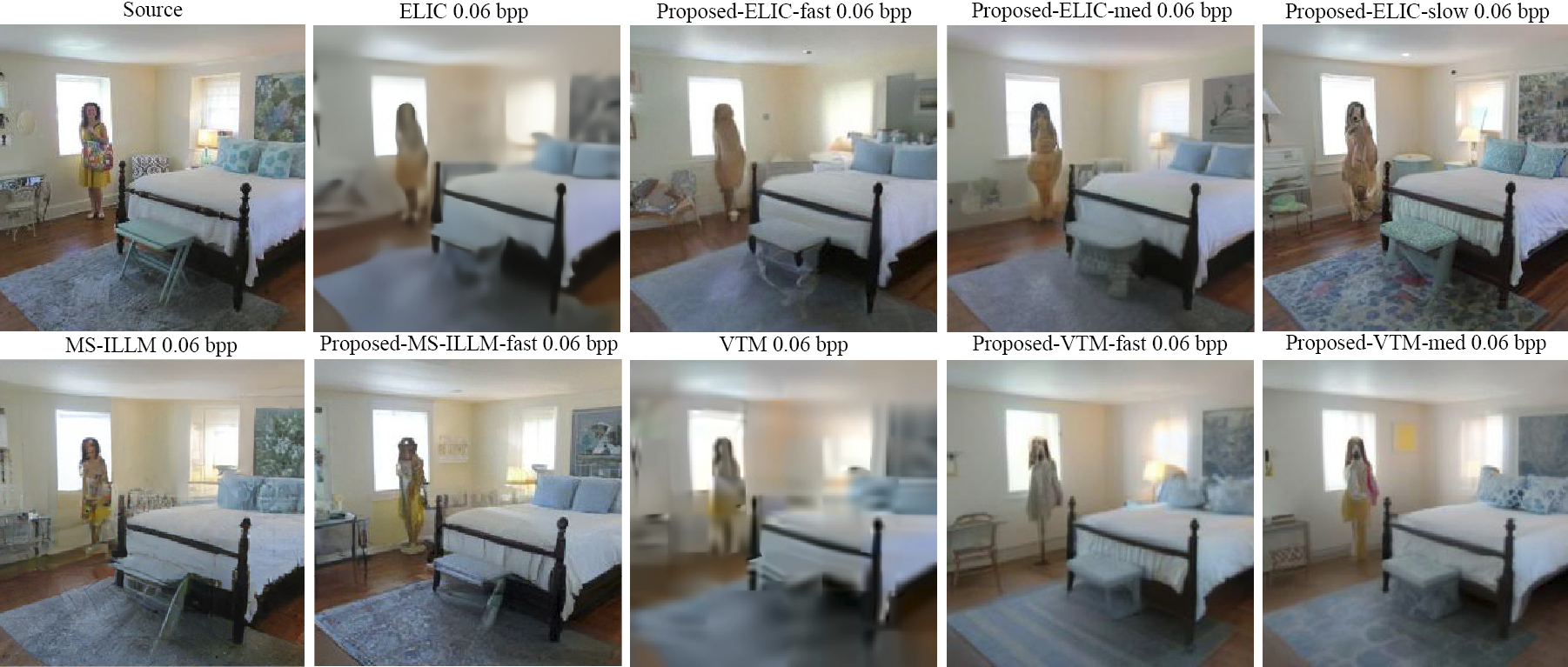}
    \includegraphics[width=\linewidth]{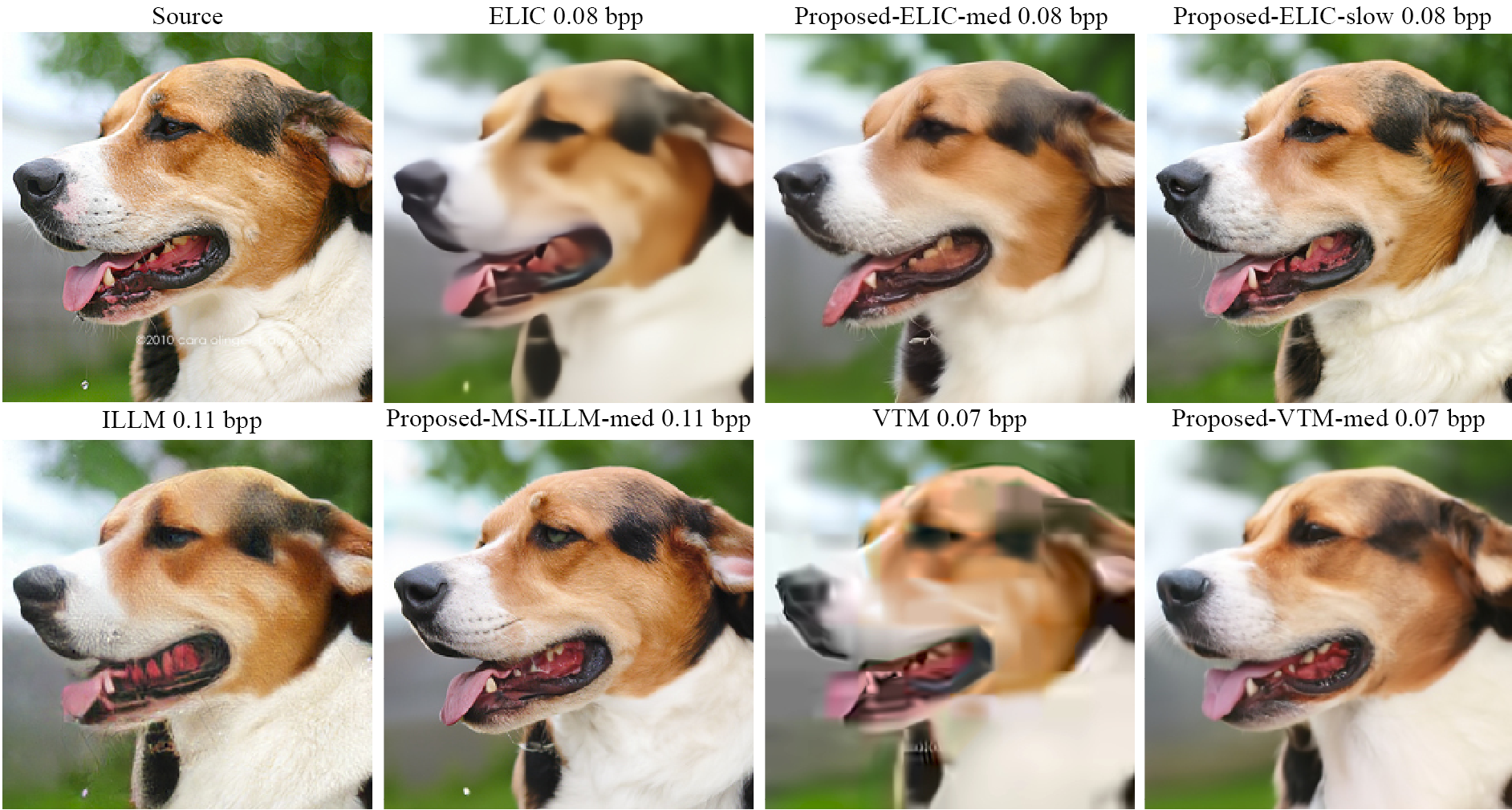}
\caption{The qualitative results of different methods.}
\label{fig:qua}
\end{figure}

We compare our proposed approach with other conditional perceptual codec and IVC with similar decoding time. For IVC, the decoding time reduction is achieved by reducing the step number of reverse diffusion process. The results are shown in Tab.~\ref{tab:rd} and Fig.~\ref{fig:qua}. First of all, in terms of overall FID, our approach is comparable to IVC, and both our approach and IVC outperform all conditional perceptual codec. However, in this $\ge 10$s decoding setting, our approach is around $\times 3-6$ times faster than IVC. In other decoding time such as $0.1$s and $0.1-10$s, our approach outperforms IVC significantly in FID. This is because the diffusion inversion \citep{Chung2022DiffusionPS} in IVC does not perform well when the step number is small. 

On the other hand, our approach uniformly improves the FID of base codec such as ELIC, VTM and MS-ILLM as decoding time increase. Our fast preset reduces the FID of all base codec significantly within $0.1$s decoding time. Our medium preset achieves even better FID in several seconds. And our slow preset achieves a FID comparable to IVC. Furthermore, the BD-PSNR of our proposed approach lies within the $-3$dB lowerbound induced by perception-distortion trade-off \citep{blau2018perception,blau2019rethinking}.

\begin{table}[thb]
\caption{Rate distortion performance on LSUN Bedroom and ImageNet. \textbf{Bold}: lowest FID. \underline{Underline}: second lowest FID.}
\label{tab:rd}
\centering
\resizebox{\linewidth}{!}{
\begin{tabular}{@{}lcccccc@{}}
\toprule
                      & \multicolumn{3}{c}{LSUN Bedroom} & \multicolumn{3}{c}{ImageNet} \\ \cmidrule(l){2-7} 
                      & Dec time $\downarrow$  & BD-FID $\downarrow$ & BD-PSNR $\uparrow$ & Dec time $\downarrow$ & BD-FID $\downarrow$ & BD-PSNR $\uparrow$ \\ \midrule
\multicolumn{7}{@{}l@{}}{\textit{MSE Codec}}                                                           \\
Hyperprior (Baseline)            &     0.02       &       0.00   &    0.00      &    0.02       &   0.00     &    0.00     \\
ELIC                  &      0.05      &    -1.56      &     1.95     &      0.05     &   -7.20     & 1.17        \\
VTM                   &      0.1      &     -3.72     &     1.90     &      0.1     &   -9.81     &  0.96       \\ \midrule
\multicolumn{7}{@{}l@{}}{\textit{Conditional Perceptual Codec}}                                        \\
HiFiC                 &       0.5     &     -14.12     &     -1.13     &     0.5      &   -33.27    &    -0.90     \\
CDC                   & 18.0 & -19.90 & -6.30 & 18.0 & -41.75 & -6.46 \\
PO-ELIC               & - & - & - &     0.05      &    -41.50    &    0.22     \\
MS-ILLM               &     0.03       &    -20.41      &     -0.87     &    0.03       &    -40.24    &    -0.55     \\ \midrule
\multicolumn{7}{@{}l@{}}{\textit{Training-free Perceptual Codec}}                                      \\ \cdashline{1-7}
\multicolumn{7}{@{}l@{}}{$\approx 0.1$s \textit{decoding}}                                      \\
IVC-ELIC      &      0.2      &     208.50     &    -20.6      &     -      &    -    &     -    \\
Proposed-ELIC (fast) &    0.1   &    -17.34     &     0.17     &   -   &   -     &     -    \\
Proposed-VTM (fast)  &     0.2       &    -16.83      &    -0.047      &    -       &    -    &     -    \\
Proposed-MS-ILLM (fast) &    0.1     &    -22.84      &     -1.05     &    -       &    -    &     -    \\ \cdashline{1-7}
\multicolumn{7}{@{}l@{}}{$0.1-10$s \textit{decoding}}                                      \\
IVC-ELIC      &      3.7      &     31.09     &     -8.16     &    2.0       &    217.00     &    -19.20     \\
Proposed-ELIC (med) &      3.2      &     -23.46     &    -0.51      &     1.7      &   -34.96     &   -1.42      \\
Proposed-VTM (med)  &      3.3      &     -23.49     &    -0.46      &     1.8      &    -35.77    &     -1.75    \\
Proposed-MS-ILLM (med) &     3.2       &     \underline{-26.54}     &     -2.32     &      1.7     &    -42.77    &   -2.79      \\ \cdashline{1-7}
\multicolumn{7}{@{}l@{}}{$\ge 10$s \textit{decoding}}                                      \\
IVC-ELIC      &      45.7      &    \textbf{-26.58}      &      -1.03    &     255.8      &   \textbf{-46.14}     &  -1.69       \\
Proposed-ELIC (slow) &     14.3       &     \textbf{-26.58}     &     -0.24     &      46.8     &    \underline{-44.72}     &    -0.67     \\ \bottomrule
\end{tabular}
}
\end{table}
\begin{figure}[thb]
\centering
    \includegraphics[width=0.75\linewidth]{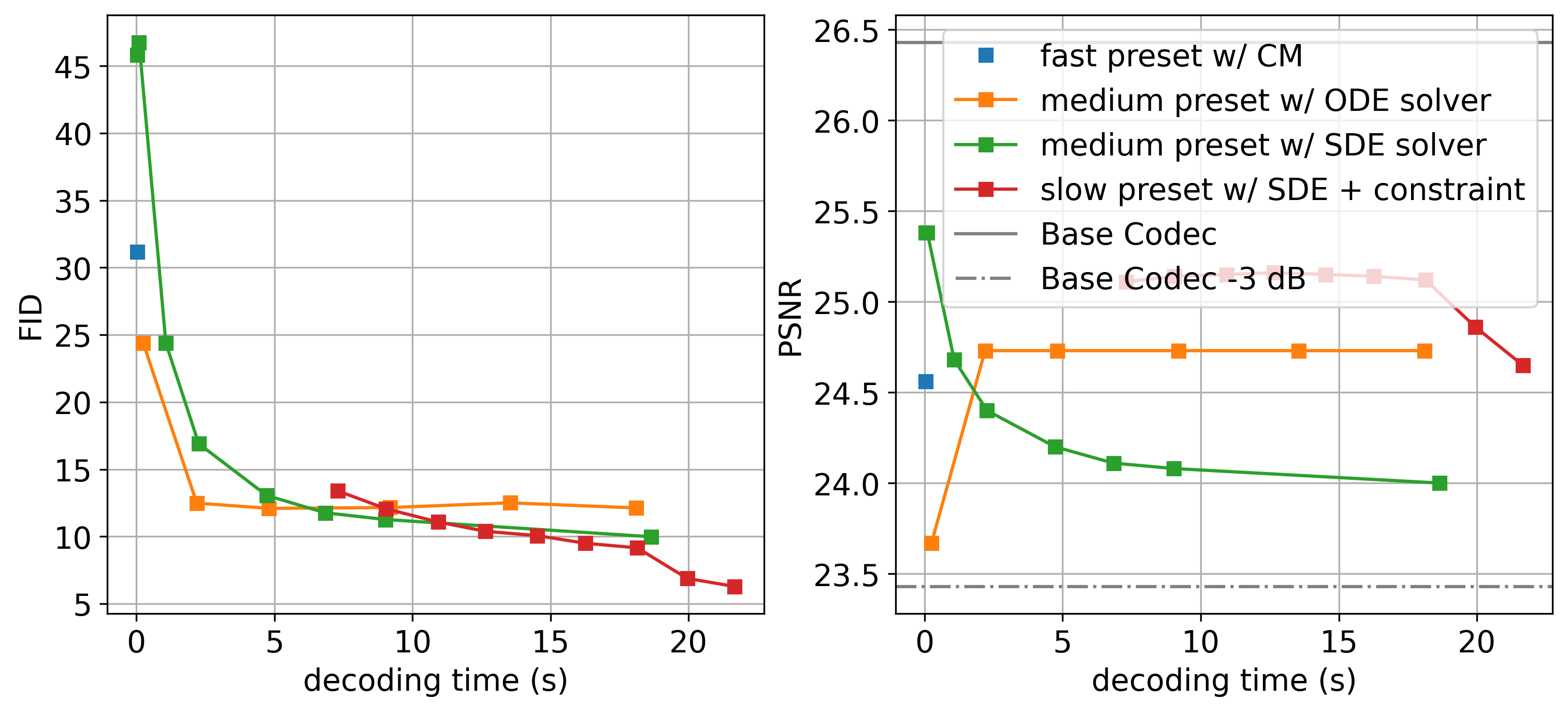}
\caption{The speed-performance trade-off of different implementations.}
\label{fig:pt}
\end{figure}
\subsection{Speed-Performance Trade-off}
We determine the different implementation of each preset in Tab.~\ref{tab:fms} by empirical results. As shown in Fig.~\ref{fig:pt}, when the decoding time is limited to $\approx 0.1$s, the only choice is consistency model, as no other approach can produce a reasonable result for $1$ step. When the decoding time is around $0.1-5$s, the ODE solver dominates the FID metric. Here, we adopt DPM solver \citep{lu2022dpm}, a fast ODE solver designed to solve PF-ODE. When the decoding limit is beyond $5$s, the SDE solver has the best FID metric. Here, we adopt the simplest Euler SDE solver. This is expected as SDE outperforms ODE when discretization error is small \citep{deveney2023closing}. But ODE solver converges faster when step number is small. When the decoding time limit is beyond $10$s, the SDE solver with diffusion inversion constraint performs the best, despite it does not perform as well as SDE solvers without constraint when decoding time is smaller.
\begin{figure}[thb]
\centering
    \includegraphics[width=\linewidth]{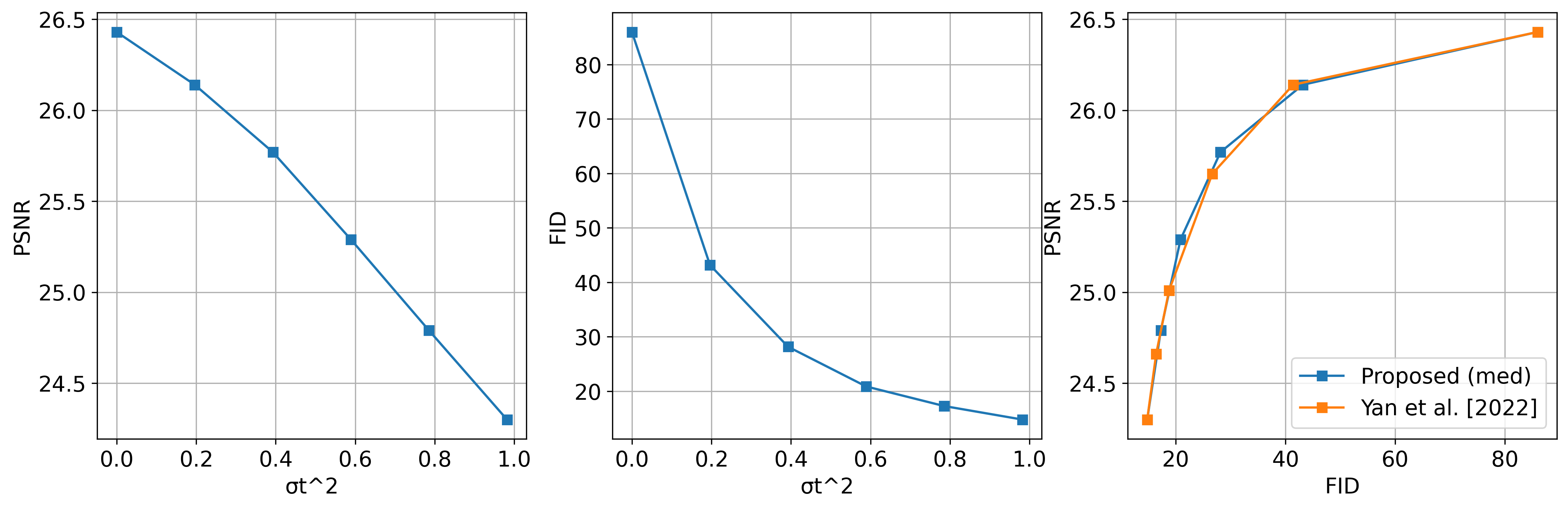}
\caption{The perception-distortion trade-off by adjusting $\sigma_t^2$.}
\label{fig:pd}
\end{figure}
\subsection{Perception-Distortion Trade-off}
When the base codec is MSE optimized, such as VTM and ELIC \citep{he2022elic}, our approach can achieve perception-distortion trade-off naturally by adjusting $\sigma_t^2$. We present this result with ELIC base codec, Bedroom dataset and medium preset. As shown in Fig.~\ref{fig:pd} and Fig.~\ref{fig:pdimg}, as $\sigma_t^2$ increases, both the PSNR and the FID drop uniformly. However, when the PSNR drop is approximately $3$dB, the FID curve starts to flatten. And this result coincides the distortion-perception trade-off theory \citep{blau2018perception, blau2019rethinking}, which states that optimal perceptual quality can be achieved at the cost of $-3$dB PSNR or $\times 2$ MSE.

Our training-free perceptual codec does not change the bitstream of base codec. Therefore, we can also achieve perception-distortion trade-off by the convex interpolation approach proposed by \citet{yan2022optimally}. They demonstrate that a simple interpolation between MSE optimized image and perceptually optimized image can achieve optimal perception-distortion trade-off when perception divergence is W2 and distortion is MSE. We compare our perception-distortion trade-off by adjusting $\sigma_t^2$ with \citet{yan2022optimally} in Fig.~\ref{fig:pd}. It is shown that our approach and \citet{yan2022optimally} have comparable performance in terms of trading-off FID and PSNR. However, our approach is simpler to implement as it only need to decode one image from a specific $\sigma_t^2$. While \citet{yan2022optimally} require decoding two images and performing an interpolation.

\section{Related Works}
\textbf{Perceptual Image Compression.} The majority of perceptual image compression adopt conditional generative model. More specifically, a conditional GAN \citep{agustsson2019generative,Mentzer2020HighFidelityGI,Agustsson2022MultiRealismIC,muckley2023improving} or diffusion model \citep{yang2022lossy,ghouse2022neural,hoogeboom2023high,careil2023towards} is trained to approximate the posterior of natural image conditioned on the bitstream. Those approaches are theoretically justified by perception-distortion trade-off \citep{blau2018perception,blau2019rethinking}.

The contemporary generative models, such as diffusion models, are hard to train. Motivated by this, \citet{Theis2022LossyCW} propose the first training-free perceptual image codec by using pre-trained VDM. More specifically, they propose to encode the trajectory of diffusion process by sample communication \citep{li2018strong}. which is intractable for practical size images. \citet{xu2024idempotence} reduce the decoding time of \citet{Theis2022LossyCW} from intractable to around 1 minute by using diffusion inversion \citep{Chung2022DiffusionPS}. However, this decoding latency remains far from practical. In this paper, we reduce the decoding time of training-free perceptual codec from $1$ min to $0.1$s, which is even comparable to those conditional generative perceptual codec. 

\textbf{Diffusion Inversion and Editing.} Diffusion Inversion are a family of algorithms that aim to solve the inverse problem with diffusion prior in a training-free manner. Most early works of diffusion inversion only fits into linear problem. \citet{Chung2022DiffusionPS} are the first to propose a diffusion inversion method for general non-linear problem. And \citet{xu2024idempotence} are the first to realize that diffusion inversion can be use to solve training-free perceptual codec. They apply the method by \citet{Chung2022DiffusionPS} to image compression and make the training-free perceptual codec depicted by \citet{Theis2022LossyCW} tractable.

\citet{meng2021sdedit} first propose to edit images by adding noise to edited image and denoise to retain perceptual quality. This paradigm has been popular in zero-shot image editing \citep{lugmayrinpainting}. Very recently, \citet{nie2024the} give this practice a theoretical justification by accurately depicting the contraction of SDE and ODE. \citet{chung2022come} also apply this approach to speed up the convergence of diffusion inversion. The resulting approach is similar to our slow preset. We are the first to realize that the approach by \citet{meng2021sdedit} can also be applied to image codec. And this realization enables $0.1$s training-free codec by using consistency model.

\section{Discussion \& Conclusion}
One limitation of our paper is that we can not scale our approach to larger images yet, because no consistency model is available for diffusion model at that scale. One possible workaround is latent consistency model \citep{luo2023latent} and latent diffusion \citep{rombach2022high}. However, latent consistency model is trained for text conditioned generation and may not be used to compress images directly. Mining text during the diffusion process can also increase decoding time \citep{Chung2023PrompttuningLD}.

To conclude, we propose a fast training-free approach that guarantees to improve the perceptual quality of any codec. Our approach decreases the decoding time from $1$ min to $0.1-10$s. Further, it can be applied to non-differentiable codec, and can be used to improve the perceptual quality of previous perceptual codec. Its perceptual quality is comparable to previous training-free perceptual codec, and is better than previous conditional generative codec.

\bibliography{main}

\clearpage

\appendix

\section{Proof of Main Results}
\label{app:pf}
\begin{proof}
For the first part, we note that $\hat{X}_t,X_t$ is a convolution of $\hat{X},X_0$ by a Gaussian kernel $\mathcal{N}(0,t^2I)$. By property of Markov chain, we have
\begin{gather}
    D_{KL}(p(\hat{X}_t)||p(X_t)) < D_{KL}(p(\hat{X})||p(X_0)).\label{eq:fmc}
\end{gather}
Next, we can adopt the Theorem 3.1 of \citet{nie2024the}, which states that for SDE,
\begin{gather}
    \forall s < t, D_{KL}(p(\hat{X}_{0|t})||p(X_{0})) = D_{KL}(p(\hat{X}_t)||p(X_t)) - \int_{0}^{t}(\frac{d\sigma^2_{\tau}}{d\tau} )D_{Fisher}(p(X_{\tau|t})||p(X_{\tau}))d\tau.\label{eq:bsde} \\
    < D_{KL}(p(\hat{X}_t)||p(X_t)).
\end{gather}
And for ODE,
\begin{gather}
    \forall s < t, D_{KL}(p(\hat{X}_{0|t})||p(X_{0})) = D_{KL}(p(\hat{X}_t)||p(X_t)).\label{eq:bode}
\end{gather}
Chaining Eq.~\ref{eq:bsde} and Eq.~\ref{eq:bode} with Eq.~\ref{eq:fmc}, we have
\begin{gather}
    D_{KL}(p(\hat{X}_{0|t})||p(X_0)) \le D_{KL}(p(\hat{X}_t)||p(X_t)) < D_{KL}(p(\hat{X})||p(X_0)).
\end{gather}
In other words, our proposed approach guarantees to improve the perceptual quality of image codec.

Next, we will show that the higher the time $t$ is, the closer the resulting KL divergence is. For diffusion process, the forward diffusion with $t>s$ can be seen as convolution $\hat{X}_s,X_s$ with Gaussian kernel $\mathcal{N}(0,(t^2-s^2)I)$. And by Markov property, we have
\begin{gather}
    D_{KL}(p(\hat{X}_t)||p(X_t)) < D_{KL}(p(\hat{X}_s)||p(X_s)).
\end{gather}
Then when $\Phi(.)$ is ODE solver, following Eq.~\ref{eq:bode}, we have
\begin{gather}
    D_{KL}(p(\hat{X}_{0|t})||p(X_0)) = D_{KL}(p(\hat{X}_t)||p(X_t)) \notag \\ < D_{KL}(p(\hat{X}_{0|s})||p(X_0)) = D_{KL}(p(\hat{X}_s)||p(X_s)).
\end{gather}
And when $\Phi(.)$ is SDE solver, following Eq.~\ref{eq:bsde}, we have
\begin{gather}
    D_{KL}(p(\hat{X}_{0|t})||p(X_0)) = D_{KL}(p(\hat{X}_t)||p(X_t)) - \int_{0}^{t}(\frac{d\sigma^2_{\tau}}{d\tau} )D_{Fisher}(p(X_{\tau|t})||p(X_{\tau}))d\tau \\ < D_{KL}(p(\hat{X}_{s})||p(X_s)) - \int_{0}^{s}(\frac{d\sigma^2_{\tau}}{d\tau} )D_{Fisher}(p(X_{\tau|t})||p(X_{\tau}))d\tau \notag \\
    = D_{KL}(p(\hat{X}_{0|s})||p(X_0)).
\end{gather}
\end{proof}

\section{Additional Experimental Setup}
\label{app:setup}
All the experiments are implemented by Pytorch, and run in a computer with AMD EPYC 7742 CPU and Nvidia A30 GPU.

\section{Additional Experimental Results}
\subsection{Additional Rate-Distortion Results}
\begin{figure}[thb]
\centering
    \includegraphics[width=\linewidth]{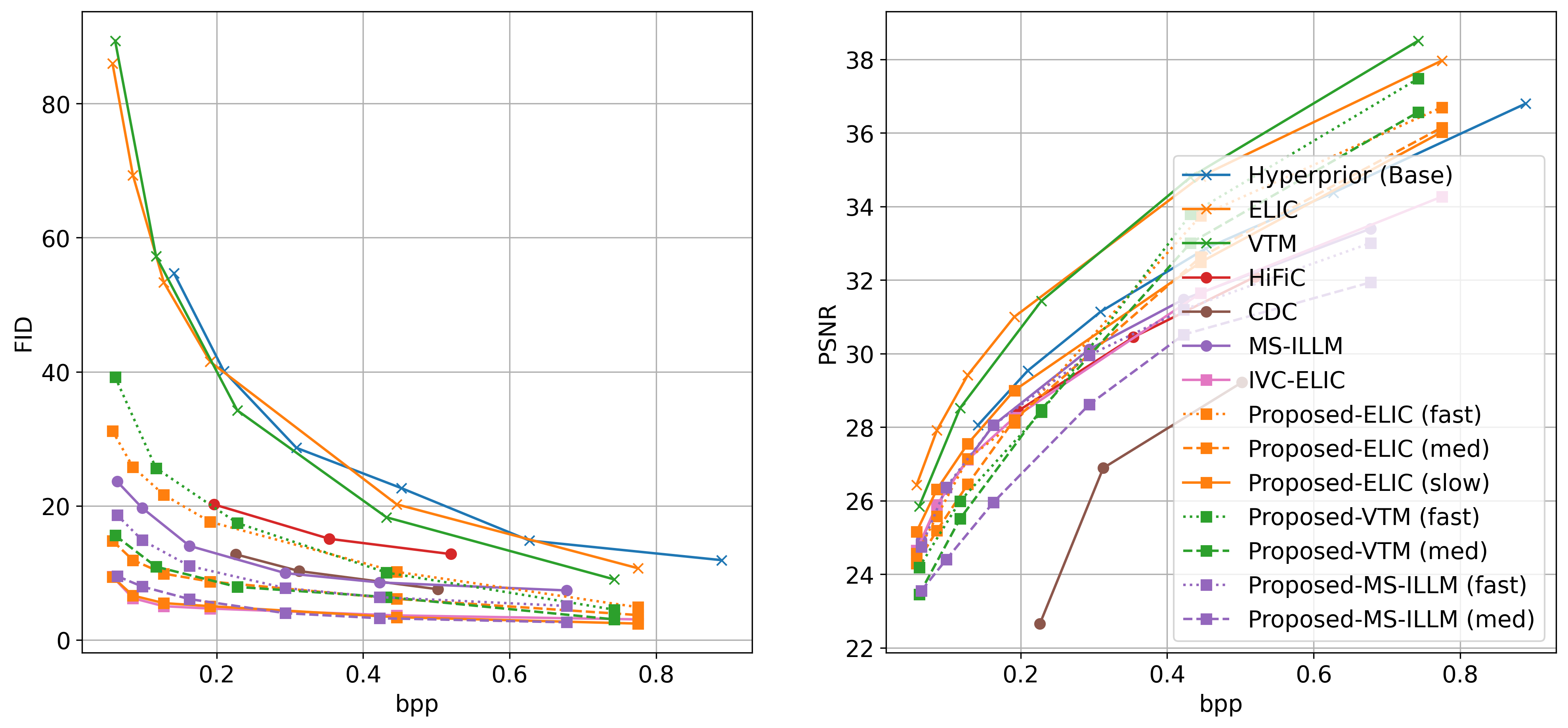}
    \includegraphics[width=\linewidth]{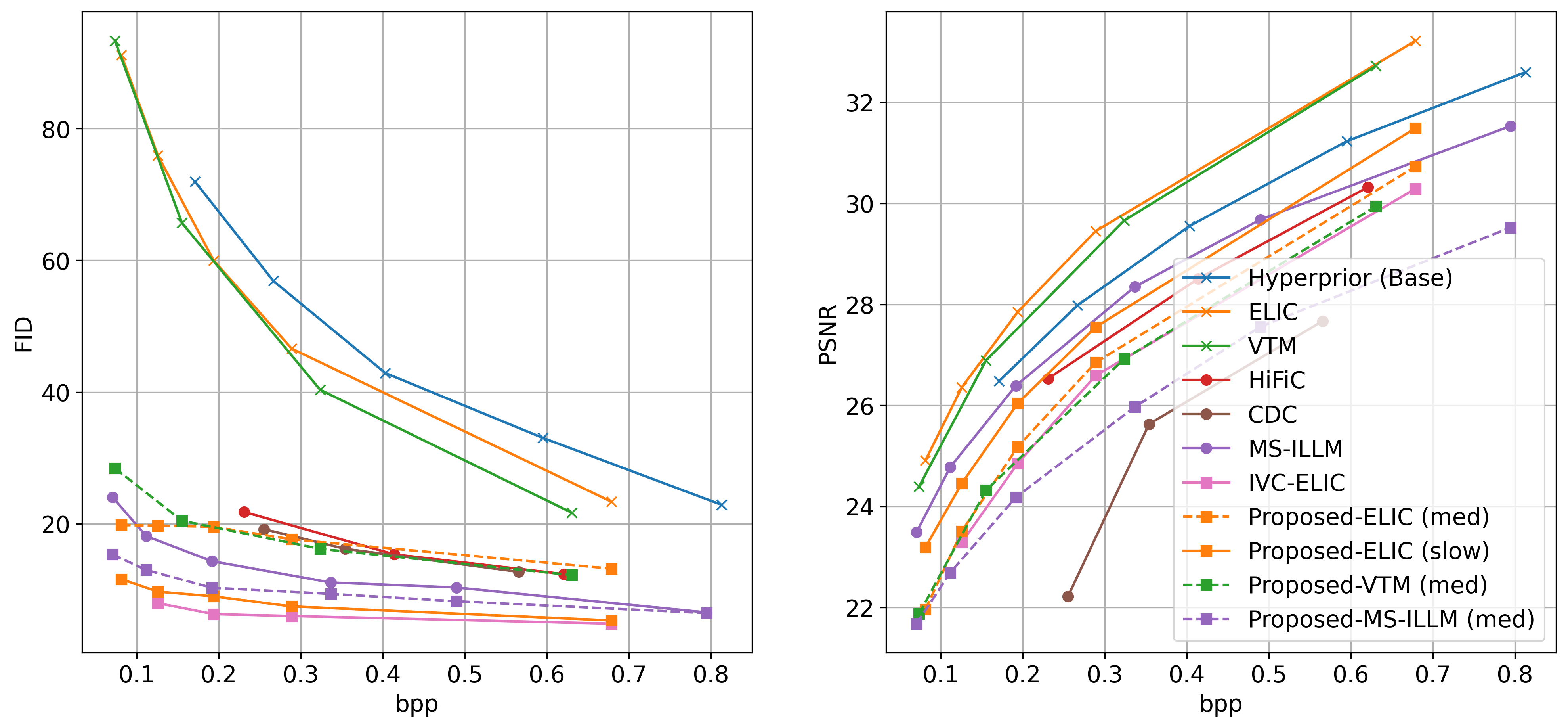}
\caption{Rate-Distortion curve of different methods.}
\label{fig:rd}
\end{figure}

We present detailed rate-distortion curve in Fig.~\ref{fig:rd}.

\subsection{Additional Results on Perception-Distortion Trade-off}
Alone with the quantitative results of perception-distortion trade-off in Fig.~\ref{fig:pd}, we present the qualitative results of perception-distortion trade-off in Fig.~\ref{fig:pdimg}. It is clearly shown that as $\sigma_t^2$ increase, the perceptual quality of the image is improved. However, it also deviates more from the MSE optimal result.

\subsection{Additional Qualitative Results}
\begin{figure}[thb]
\centering
    \includegraphics[width=\linewidth]{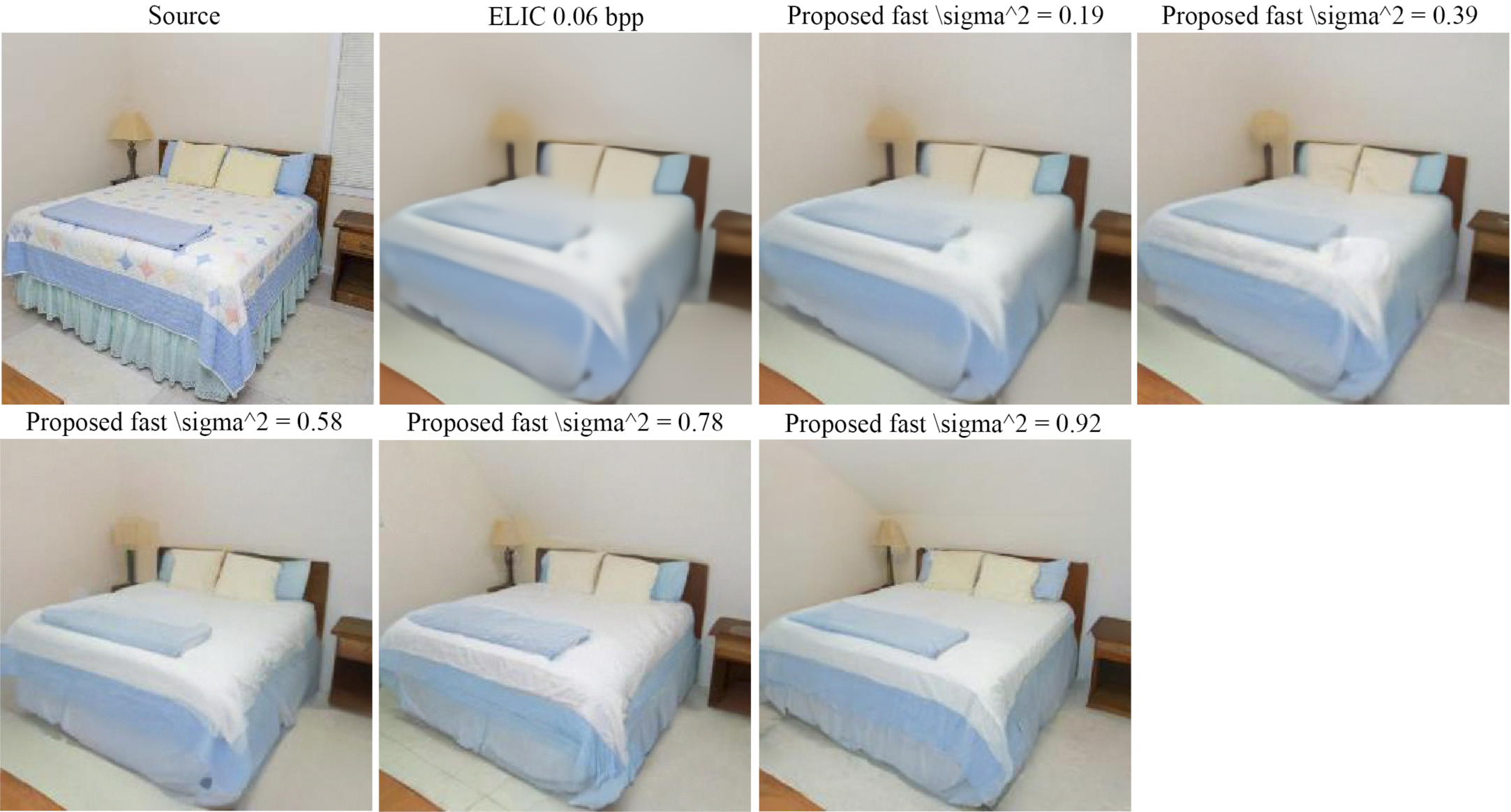}
\caption{Qualitative results of perception-distortion trade-off.}
\label{fig:pdimg}
\end{figure}

Alone with the quantitative results of perception-distortion trade-off in Fig.~\ref{fig:pd}, we present the qualitative results of perception-distortion trade-off in Fig.~\ref{fig:pdimg}. It is clearly shown that as $\sigma_t^2$ increase, the perceptual quality of the image is improved. However, it also deviates more from the MSE optimal result.

\begin{figure}[thb]
\centering
    \includegraphics[width=\linewidth]{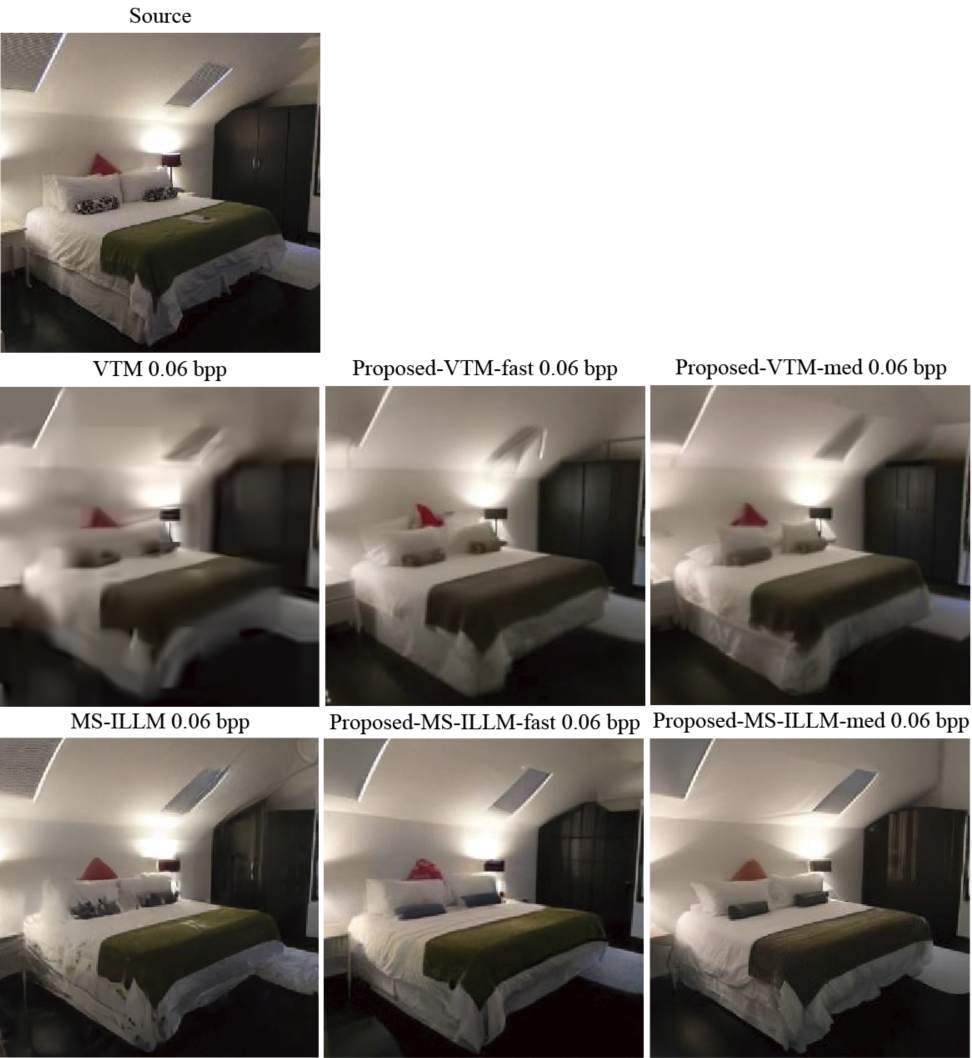}
\caption{Additional qualitative result on LSUN Bedroom dataset.}
\label{fig:lsunqual2}
\end{figure}

\begin{figure}[thb]
\centering
    \includegraphics[width=\linewidth]{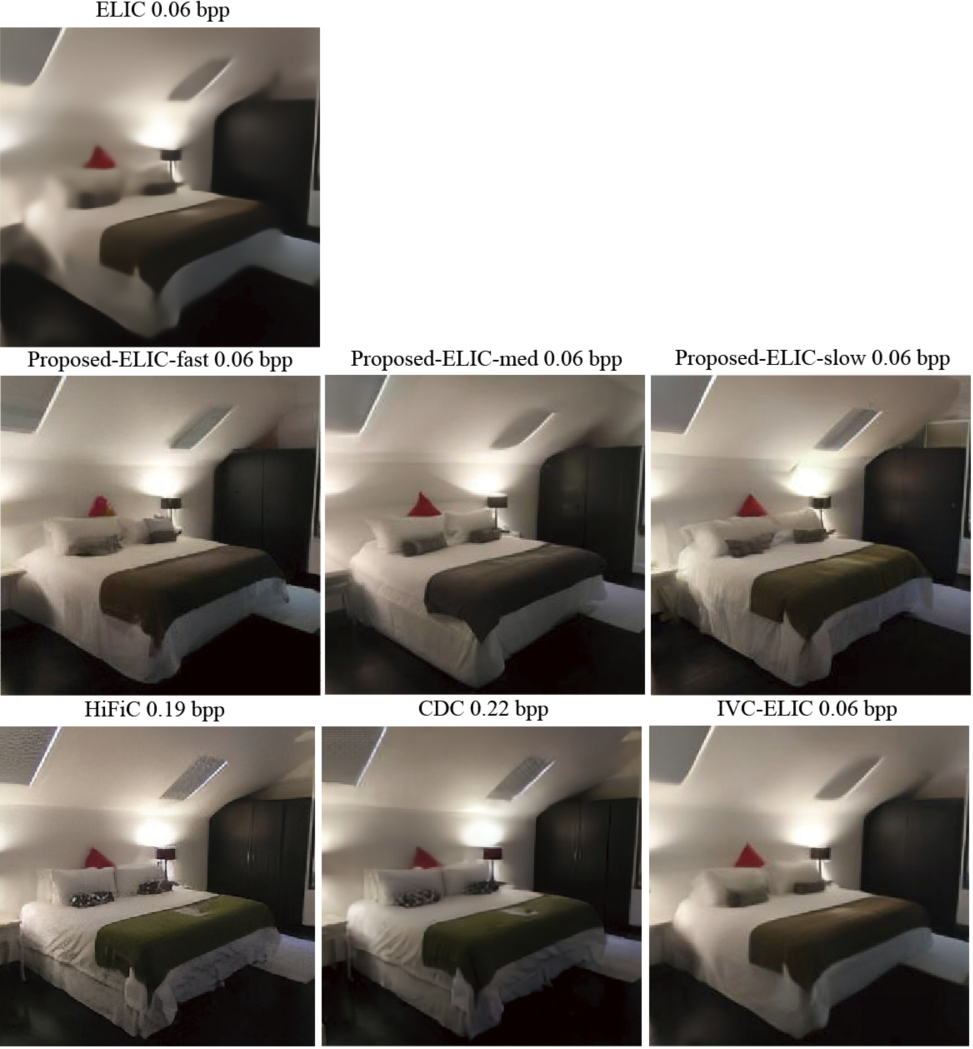}
\caption{Additional qualitative result on LSUN Bedroom dataset.}
\label{fig:lsunqual3}
\end{figure}

\begin{figure}[thb]
\centering
    \includegraphics[width=\linewidth]{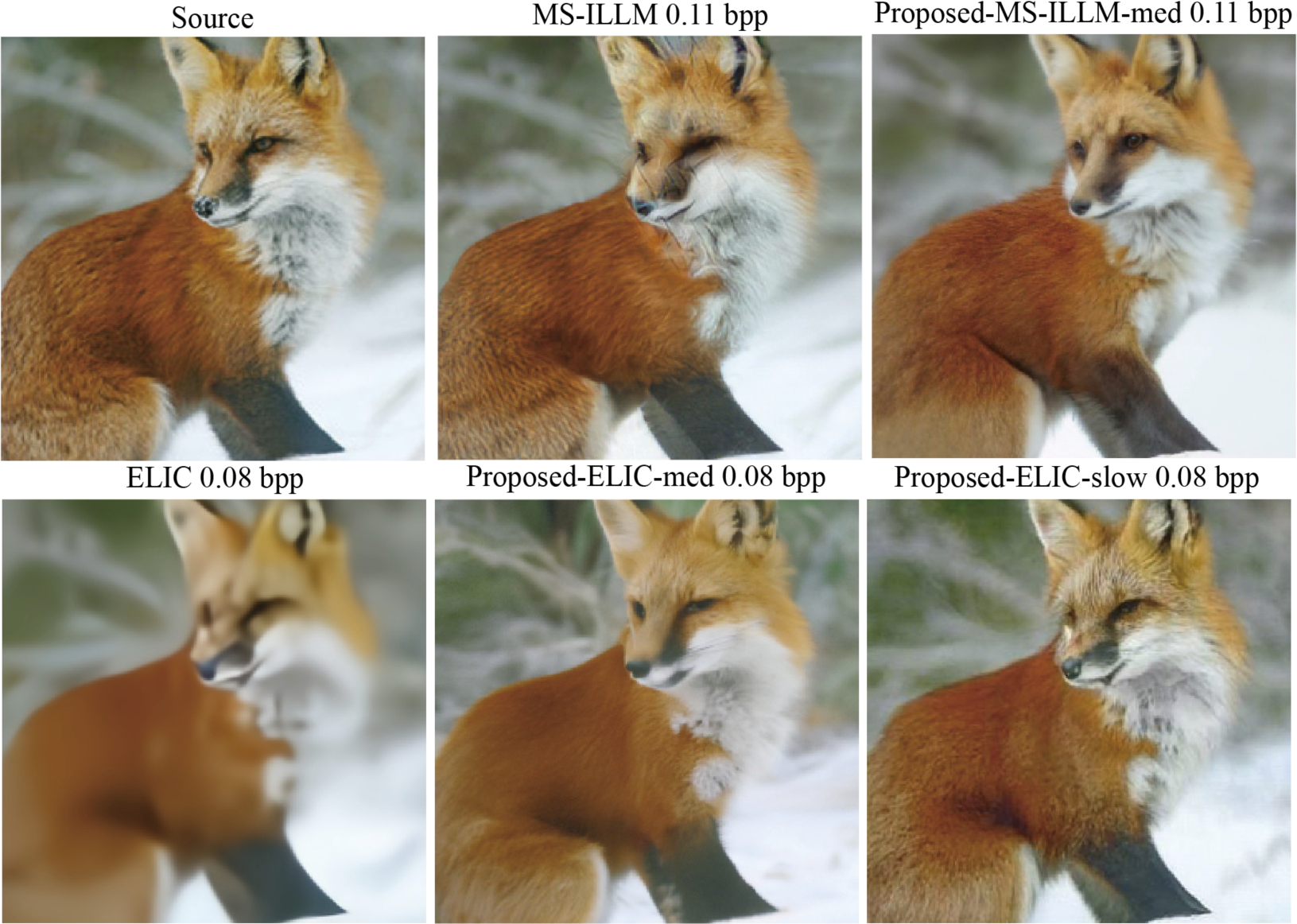}
    \includegraphics[width=\linewidth]{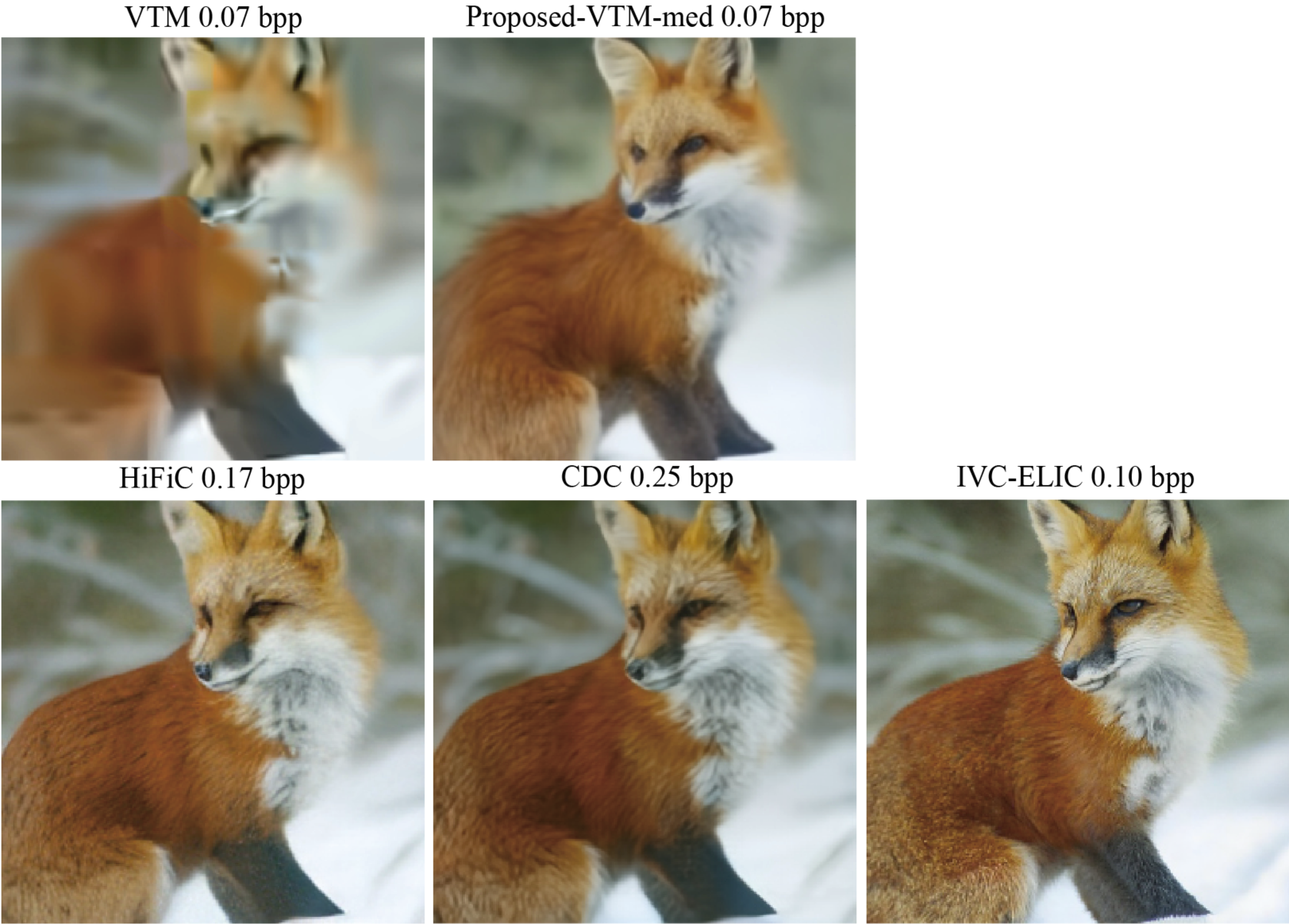}
\caption{Additional qualitative result on ImageNet dataset.}
\label{fig:imgqual3}
\end{figure}

\begin{figure}[thb]
\centering
    \includegraphics[width=\linewidth]{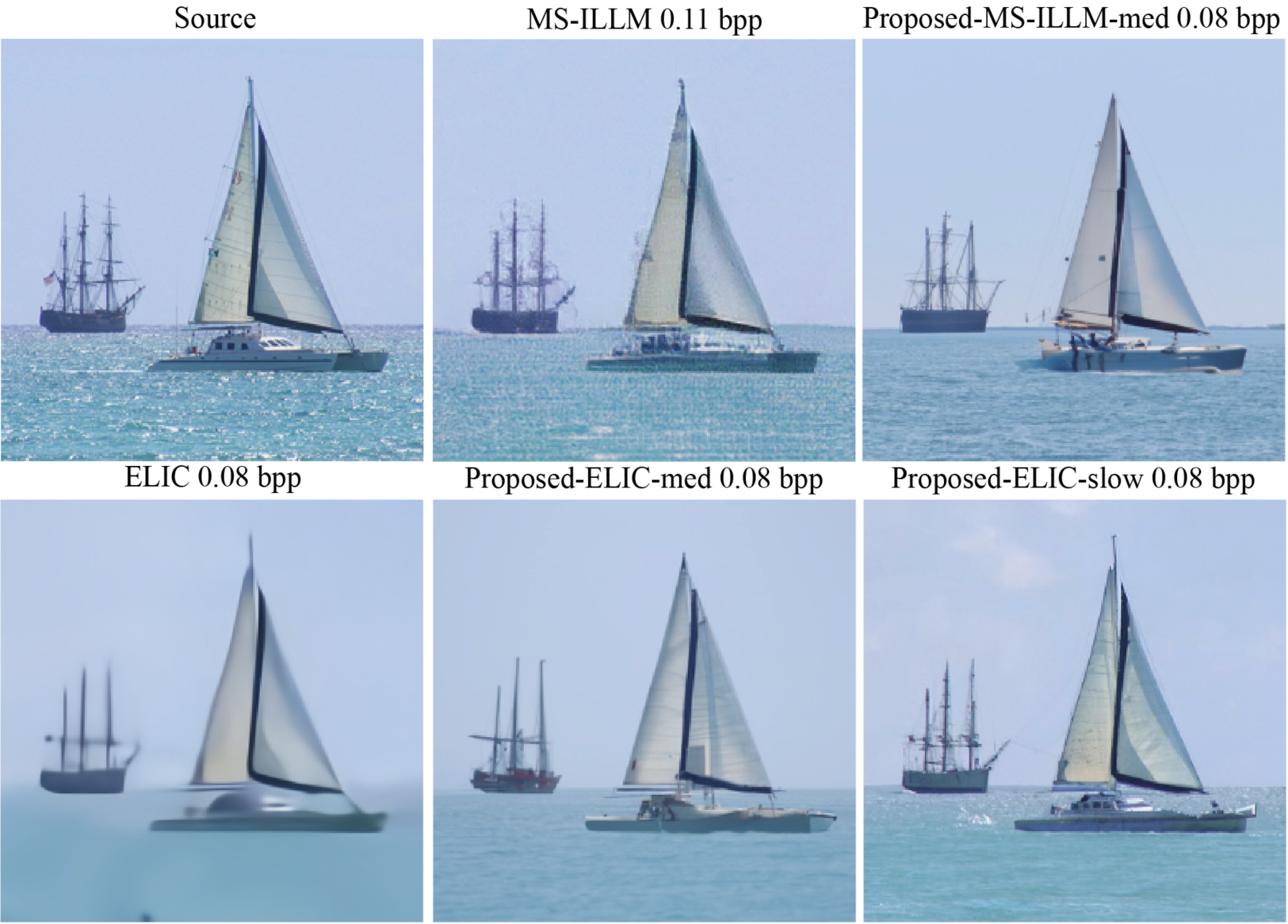}
    \includegraphics[width=\linewidth]{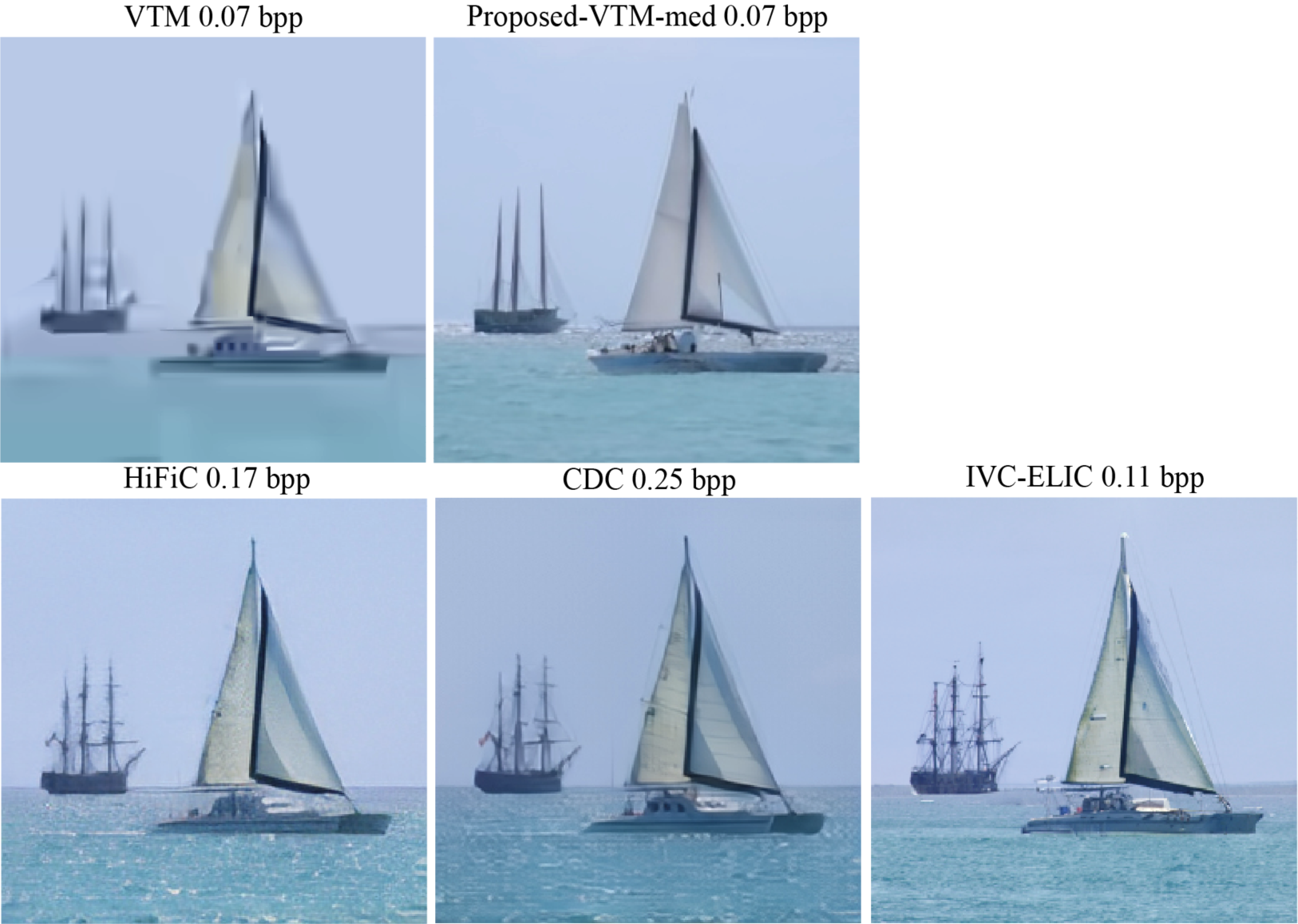}
\caption{Additional qualitative result on ImageNet dataset.}
\label{fig:imgqual2}
\end{figure}

We show additional qualitative result on LSUN Bedroom in Fig.~\ref{fig:lsunqual2} and Fig.~\ref{fig:lsunqual3}, and additional qualitative result on ImageNet in Fig.~\ref{fig:imgqual3} and Fig.~\ref{fig:imgqual2}.

\subsection{Failure Case}
\begin{figure}[thb]
\centering
    \includegraphics[width=\linewidth]{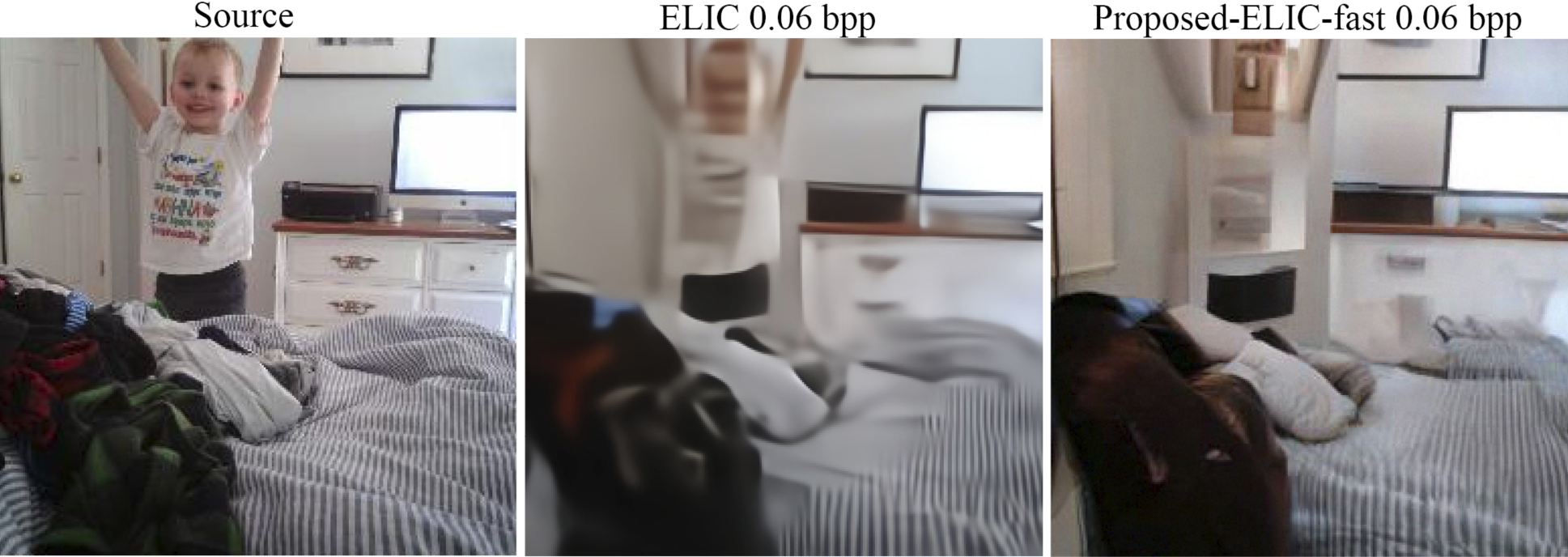}
\caption{A typical failure case with out of distribution input.}
\label{fig:fa}
\end{figure}

Our method is not able to handle out-of-distribution images. For example, a typical failure case is shown in Fig.~\ref{fig:fa}. The base diffusion model is EDM trained with LSUN Bedroom. And obviously our proposed approach can not handle human in the scene. The little boy is transformed into some furniture after our perceptual enhancement with fast preset.

\section{Additional Discussion}
\subsection{Reproducibility Statement}
The proof of theoretical results are shown in Appendix.~\ref{app:pf}. All two datasets are publicly available. We provide detailed experimental setup in paper and appendix. Besides, we provide source code in supplementary material.

\subsection{Broader Impact}

The approach proposed in this paper saves both training time and inference time of perceptual codec. This has positive social values including reducing the carbon footprint of producing and deploying codec. On the other hand, as our codec is generative, potential negative impact may be inherited from other generative model. 

\clearpage

\end{document}